\documentclass[12pt]{article}
\usepackage{epsfig}

\parskip=\medskipamount    
\renewcommand{\thesection}{\arabic{section}}

\catcode`@=11
\@addtoreset{equation}{section}
\@addtoreset{equation}{subsection}
\def\theequation{\ifnum\value{section}=0 \arabic{equation}\ignorespaces
\else \ifnum\value{section}=-1 A.\arabic{equation}\ignorespaces
\else \ifnum\value{subsection}=0 \thesection.\arabic{equation}\ignorespaces
\else \thesection.\arabic{subsection}.\arabic{equation}\ignorespaces
              \fi
                        \fi
                   \fi}

{\catcode`\'=\active \def'{{}^\bgroup\prim@s}}
\catcode`@=12

\newcommand{\bq}{\begin{equation}}
\newcommand{\be}{\begin{equation}} 
\newcommand{\fq}{\end{equation}}
\newcommand{\ee}{\end{equation}}
\newcommand{\bqr}{\begin{eqnarray}}
\newcommand{\beqs}{\begin{eqnarray}} 
\newcommand{\fqr}{\end{eqnarray}}
\newcommand{\eeqs}{\end{eqnarray}}

\def\pa{\partial}

\def\bop#1{\setbox0=\hbox{$#1M$}\mkern1.5mu
  \vbox{\hrule height0pt depth.04\ht0
    \hbox{\vrule width.04\ht0 height.9\ht0 \kern.9\ht0
    \vrule width.04\ht0}\hrule height.04\ht0}\mkern1.5mu}
\def\Box{{\mathpalette\bop{}}}                    

\begin{document}
\thispagestyle{empty}

\def\Prop{\Triangle} 
\def\Prod{\prod} 

\begin{flushright} 
\begin{tabular}{l} 
UCLA-02-TEP-26\\
hep-th/0209088 \\ 
\end{tabular} 
\end{flushright}  

\vskip .3in 
\begin{center} 

{\Large\bf  N=4 Supersymmetric Gauge Theory in the Derivative Expansion} 

\vskip .3in 

{\bf Gordon Chalmers} 
\\[5mm] 
{\em Department of Physics and Astronomy \\ 
University of California at Los Angeles \\ 
Los Angeles, CA  90025-1547 } \\  

{e-mail: chalmers@physics.ucla.edu}  

\vskip .5in minus .2in 

{\bf Abstract}   
\end{center} 

Maximally supersymmetric gauge theories have experienced renewed interest 
due to the AdS/CFT correspondence and its conjectured S-duality.  These gauge 
theories possess a large amount of symmetry and have quasi-integrable 
properties. 
We derive the amplitudes in the derivative expansion of the spontaneously 
broken examples and perform all loop integrations.  The S-matrix is found via 
an algebraic recursion and at each order is SL(2,Z) invariant.

\setcounter{page}{0}
\newpage 
\setcounter{footnote}{0} 

\section{Introduction} 

Supersymmetric gauge theories with four conserved supersymmetries 
is conjectured to have a non-perturbative self-equivalence \cite{Montonen:sn, 
Witten:mh} and 
has been studied for many reasons.  These theories have the most 
symmetry of gauge theories at arbitrary couplings, including 
conformal invariance.  The well-known holographic description 
of IIB string theory via $N=4$ gauge theory \cite{Maldacena:1997re, 
Gubser:1998bc, Witten:1998qj,Aharony:1999ti}, or vice versa, generates
information in the latter at the non-perturbative corner and 
is a well-defined example of the gravity/gauge correspondence.  
The construction of the S-matrix in this work has manifest 
S-duality and unitarity at each order in the expansion.\footnote{The 
S-duality 
invariance within the AdS/CFT correspondence was explored 
to all orders in \cite{Chalmers:2000vq,Chalmers:2000zg}, and leads 
to the composite operator 
correlations in the gauge theory.  It is possible to translate 
information from amplitudes to composite operator correlations.}

Spontaneously broken $N=4$ supersymmetric gauge theories have a 
mass parameter (and a BPS self-dual set of states) and an 
expansion in this variable, $\Box/m^2$, is readily available. 
S-duality of $N=4$ gauge theory requires that the S-matrix of 
gauge bosons and composite operators be invariant via, 

\bqr 
\tau \rightarrow {a\tau+b\over c\tau+d} \ , 
\fqr 
with the coupling, 
\bqr 
\tau={\theta\over 2\pi} + i {4\pi\over g^2} \ . 
\fqr 
This structure, together with the perturbative coupling dependence  
and instantons requires that the correlations be built from a classified 
set of automorphic functions, Eisenstien series and cusp form(s); for a 
review see \cite{Obers:1999um}.  One 
interesting feature of S-duality is that knowledge of the 
perturbative scattering is enough to construct the instanton series; 
together, these two series are redundant.  S-duality also might 
lead to exact solutions for correlators.  This approach has 
manifest S-duality in the expansion, and is non-perturbative.

Following along the lines of the recursion formulae in scalar 
field theory, we now consider amplitudes and correlators in 
in the broken gauge theory.  The tensor manipulations and integrals 
have already been done in \cite{scalar}.  The primary 
difference is that $N=4$ gauge theory contains a conjectured 
S-duality, and this symmetry of the amplitudes helps control the 
coupling structure.  $N=8$ was previously examined in the derivative 
expansion \cite{Chalmers:2000ks} and new insights regarding the 
finiteness was found; $N=8$ contains an $SO(8)$ $N=4$ subtheory.  

\section{Couplings and Modular Invariance}

S-duality requires the quantum generating functional of the 
S-matrix  (gauge bosons) and correlations to be invariant under 
$SL(2,z)$ transformations.  The lagrangian is, 

\bqr 
{\cal L} = \int d^4x ~ {\rm Tr} \left( F^2 + \psi \nabla \psi + 
\phi \Box \phi + [ \phi^i,\phi^j ]^2 \right) \ , 
\fqr 
with gauge group G, e.g. SU(N).  The toroidal modular parameter in 
general is a matrix, 

\bqr 
\tau_{ij} = {\theta\over 2 \pi} + {4\pi i\over g^2} \vert_{ij} \ , 
\fqr 
and for simplicity we take the gauge group to be $SU(2)$.  The complications 
from enlarging $SU(2)\rightarrow U(1)$ to $SU(N)\rightarrow U(1)^{N-1}$ is 
in the modular functions and trace structures involved in the sewing.  

We examine the gauge boson scattering through the quantum 
generating functional of the S-matrix.  This functional is built from 
the operators, 

\bqr 
{\rm Tr} F^n, ~{\rm Tr} F^n {\rm Tr} F^m, ~\ldots 
\fqr 
together with covariant derivatives.  Label the operators by the
dimension, ${\cal O}_{(d_i)}^i$; the dimensionality is labeled with 
the expectation value of a scalar field giving the correct dimensions, 

\bqr 
\langle\phi\rangle^{4-d_i} {\cal O}^i_{(d_i)} \ . 
\fqr 
Because of S-duality the terms in the quantum generating functional 
must be modular invariant functions.  These have been classified 
\cite{Chalmers:2001kx}, and 
separable into normalizable and non-normalizable ones, and convergent 
versus non-convergent.  The former condition is not relevant because it 
is tantamount to integrating over the coupling constants over the 
inequivalent 
vacua.  The nonconvergent modular forms could be zeta function regularized; 
however, this results in a modular anomaly (an example is produced below).  
Any automorphic function may be represented as an infinite sum of the 
independent ones for complex values of $s$, 

\bqr
E_s^{(q,-q)}= \sum_{p,q}  {\tau_2^s\over (p\tau+q)^{s-q} (p\bar\tau+q)^{s+q}}
\fqr 
They are divergent for ${\rm Re} s<1$.  The scattering of gravitons and the 
bounds on the planar limit of $N=4$ strong coupling correlators indicate that 
only half-integral or integral values of $s$ enter into the description of 
the scattering.  For non-vanishing $q$ these functions transform with weight 
$q$.  They satisfy the $SL(2,Z)$ invariant Laplacian, and potentially leads 
to differential relations in the terms of the S-matrix expansion. 

The set of functions we consider consist of the ring, 

\bqr 
\prod E_{s_i}^{(q_i,-q_i)}  
\fqr 
with $\sum s_i=s=n/2$, $n$ an integer, and $\sum q_i=0$ with 
$\vert q_i\vert \leq s_i$.  The amplitudes containing the fermions have 
non-zero 
$q$ ($N_\psi = q/2$).  

In $k$-space, the amplitude for four gauge bosons is an expansion, 

\bqr 
\sum {\rm Tr} \nabla^{2n} F^4 h_n(\tau,\bar\tau) 
\fqr 
with the derivatives distributed among the four field strengths 
(the color structures are implied, and multi-trace structures are 
found by taking a product of lower trace ones).  The general term 
is, 

\bqr 
\Prod_{i=1}^{p^A} \Prod_{j=1}^{n_i^A} \pa_{\mu_{\sigma'(i,j)}} 
  \Prod^{m_i^A} A_{\mu_{\sigma(i,j)}} 
\Prod_{i=1}^{p^\phi} \Prod_{j=1}^{n_i^\phi} \pa_{\mu_{\sigma'(i,j)}}
\fqr 
\bqr 
  \Prod^{m_i^\phi} \phi_{a_{\sigma(i,j)}} 
\Prod_{i=1}^{p^\psi} \Prod_{j=1}^{n_i^\psi} \pa_{\mu_{\sigma'(i,j)}}
  \Prod^{m_i^\psi} \psi_{\alpha_{\sigma(i,j)}} 
\fqr 
and may be grouped into gauge invariant operators, in the functional, 

\bqr 
\Prod_{i=1}^{p^A} \Prod_{j=1}^{n_i^A} \nabla_{\mu_{\sigma'(i,j)}}
  \Prod^{m_i^A} A_{\mu_{\sigma(i,j)}\nu_{\tilde\sigma(i,j)}}
\Prod_{i=1}^{p^\phi} \left( \Prod_{j=1}^{n_i^\phi} 
\nabla_{\mu_{\sigma'(i,j)}} \right) 
\fqr 
\bqr 
  \Prod^{m_i^\phi} \phi_{a_{\sigma(i,j)}}
\Prod_{i=1}^{p^\psi} \left( 
\Prod_{j=1}^{n_i^\psi} \nabla_{\mu_{\sigma'(i,j)}} \right) 
  \Prod^{m_i^\psi} \psi_{\alpha_{\sigma(i,j)}}  
\fqr 
with the color indices implied.  Unitarity is built into the sewing, and 
may also be seen via expanding the usual Feynman diagrams at small 
$k^2/m^2$.  In order to generate the unitarity cuts of the massive modes, 
the gauge multiplet and the dyons, one has to resum the terms in the 
derivative expansion.  Alternatively, from the expected cuts in the 
amplitudes, we may impose more conditions on coefficients beyond 
the recursion.  

The prefactors $h$ are invariant, and model the coupling dependence,  

\bqr 
h({a\tau + b\over c\tau + d},{a\bar\tau + b\over c\bar\tau + d})
= h(\tau,\bar\tau) \ . 
\fqr 
Our task is to find the ring of functions that enter into the $h$ 
functions and then to utilize the sewing to build the S-matrix.  
In this manner, unlike the $\phi^n$ theory, the coupling structure 
in $\tau$ is simplified and also allows for a determination of the 
complete set of instanton corrections to the amplitudes!  One of the 
two series, perturbative and non-perturbative, is redundant.  The 
perturbative derivative series is invariant under S-duality.  In agreement 
with the scattering functions, we examine functions $h_k$ in the ring 
with $s=m/2$, an integer or half-integer (this will be elaborated further).  
The Eisenstein functions admit the representation, 

\bqr 
E_s^{(q,-q)}= \sum_{p,q} 
{\tau_2^s\over (p\tau+q)^{s-q} (p\bar\tau+q)^{s+q}}
\fqr 
and expansion, found via a Possion resummation, 

\bqr
E_s^{(q,-q)} = 2\zeta(s) \tau_2^s + \alpha_s \tau_2^{1-s} + 
  {\cal O}(e^{-\tau_2}) \ .
\fqr 
The power terms are associated with perturbation theory, and the 
exponential series the instanton corrections.   The ring 
$S_{n/2}^{q}$ of modular invariant functions is the set, 

\bqr 
\left\{ \prod_{j=1}^p E_{s_j}^{(q_j,-q_j)} \right\} 
\fqr 
with $\sum_j s_j\leq n$ and $\sum q_j=q$.  These functions converge 
individually for $s>1$.   The regularized versions could be utilized, 
for example, ${\bar E}_1 = \ln \tau_2\eta(\tau,\bar\tau)$, but they possess 
a modular anomaly as a result.  The basis was already commented on, 
and follows from examining compatibility with the perturbative coupling 
series.  

A n-gauge boson scattering amplitude has the coupling series, 

\bqr 
\left(g^2\right)^{n/2-1}, \left(g^2\right)^{n/2}, \ldots,
\left(g^2\right)^{n/2-2+n_{max}/2} \ . 
\fqr 
We have labeled the maximum coupling with the integer $n_{max}$, as 
the ring of functions related to this expansion always truncate.  It is 
impossible to generate an infinite series in $g^2$ with $s\leq n/2$.   
The series of terms may be regrouped into, 

\bqr 
g^{n-1} \left( g^2\right)^{n_{max}/2} 
\left[ \left( {1\over g^2}\right)^{n_{max}/2}, \ldots, 
\left({1\over g^2}\right)^{-n_{max}/2+1} \right] 
= g^{n-1} \left( g^2\right)^{n_{max}/2} h_n(\tau,\bar\tau) 
\fqr 
with $k=n_{max}$, with $n_{max}=n+2$; there are no inverse
powers of $g$ in perturbation theory.  The anomalous, from the 
modular point of view, front factor may be removed with a field 
redefinition, 

\bqr 
A\rightarrow g^{-2} A  \qquad  x\rightarrow g x \ . 
\fqr 
Then the Lagrangian is, 

\bqr 
\int d^4x {1\over g^2} ~ {\rm Tr} 
 \left( \partial A + {1\over g^2} A^2\right)^2 
\fqr 
and the $g^2$ factor in front is absorbed by the implied metric $\sqrt g$
in the coordinate redefinition.  The coordinate redefinition is a slight 
improvement on the statement of S-duality of $N=4$ super gauge theory, 
similar to Einstein frame in IIB superstring theory.  

The amplitudes have the expanded form by varying the quantum generating 
functional, 

\bqr 
\langle A(k_1) \ldots A(k_n)\rangle = \sum_q h_q^{(n)} (\tau,\bar\tau)
  f_q(k_1,\ldots,k_n) 
\fqr 
with $h_q$ elements in the finite ring $S_{n/2+1}^{(0)}$.  The form is 
similar 
for the fermions, and $n+2$ is changed in general to $n_A+n_\phi+n_\psi/2 
+2$.  Via the expansion of the modular  functionis, the maximum loop 
is $n_{max}-1$ for the $n$-point function.

The four-point function in massless $N=4$ has the property that the 
$A^4$ term receives corrections at tree-level and one-loop only.  In 
the spontaneously broken theory the modular ring has 
$n_{max}=6$, $n/2+1\rightarrow 4$.  This corresponds to a truncation 
of the $A^4$ after five loops.  The relevant ring contains, 

\bqr 
E_3, E_{3/2}^2, \vert E_{3/2}^{(1,-1)}\vert^2 \ , E_2 
\fqr 
\bqr 
\vert E_{3/2}^{-1/2,1/2}\vert^2 ,
\vert E_{3/2}^{-3/2,1/2}\vert^2 \ , \vert E_{3/2}^{+3/2,-3/2}\vert^2
\fqr 
where only the convergent ones are included.  These separate functions 
leaves unknown parameters, which are determined by the derivative expansion, 
up to five loops if all functions 
are included.  This is simpler than an infinite series, as the four-point 
with no derivatives has only six functions.  The structure is 
similar at higher order in derivatives and number of gauge bosons, and 
the instanton corrections are determined by the perturbative terms.  

\section{Sewing and Amplitudes} 

The sewing relation, as presented in $\phi^n$ theory, may be used
to generate all of the coefficients in the derivative expansion.  
The approach is the same as in the scalar theory, with the added 
complication of many tensors.  However, all loop integrals may be 
performed and the amplitudes are generated via algebraic computations.  
The parameter $L$ counts $L-1$ loops in these diagrams, and for simplicity 
we keep the parameter as $L$.  
We have to sum over all the vertices, with $n_A$, $n_\phi$, and $n_\psi$ 
lines, together with multi-derivatives acting on them.  The general 
vertex is listed in the previous section.  

The recursion is illustrated in figure (1) of \cite{scalar}.  The left 
hand side takes the 
form, with $m_l+m_r=n$-point (numbers of lines to the left and right 
of the quantum vertex),  

\bqr
\sum_L  \left[ \Prod_{i=1}^{m_l} \left( \Prod_{j=1}^{m_i^\pa} 
    \pa_{\mu_{\sigma(i,j)}} A_{\mu_i}(k_i) \right) \right]  
\left[ \Prod_{i=m_l+1}^{m_r} \left( \Prod_{j=1}^{{\tilde m}_i^\pa}
       \pa_{\mu_{{\tilde\sigma}(i,j)}} A_{\nu_i}(k_i) \right) \right] 
\fqr 
\bqr 
\langle \left[ \Prod_{i=1}^L \left( \Prod_{j=1}^{m_i^\pa}
    \pa_{\mu_{\sigma'(i,j)}} \Prod_{j=1}^{m_i^A} A_{\mu_i}
        \right) \right] 
    \left[ \Prod_{i=1}^L \left( \Prod_{j=1}^{{\tilde m}_i^\pa}
  \pa_{\mu_{{\tilde\sigma}'(i,j)}} \Prod_{j=1}^{{\tilde m}_i^A}
        A_{\tilde\mu_i} \right) \right] 
\rangle \times   t^{\mu_\sigma,\mu_i,m_i^\pa} 
t^{\tilde\mu_\sigma,{\tilde\mu}_i,{\tilde m}_i^\pa}  
\fqr 
\bqr 
+ {\rm permutations} \ , 
\fqr 
with $\sum m_i^A=\sum {\tilde m}_i^A=L$; the $\tau$ is within the 
tensor $t$, as well as the coefficients of the derivative structure.  
We have to sum over all possible combinations, as iterated from 
lower to higher orders in derivatives and numbers of external 
states.  Internal fermions and scalars must be included in the 
sum.  

The right-hand side of the equation contains the quantum vertices, 

\bqr 
t^{\mu_\sigma,{\mu}_i,m_i^\pa}
\Prod_{i=1}^{p^A} \Prod_{j=1}^{n_i^A} \pa_{\mu_{\sigma'(i,j)}} 
  \Prod^{m_i^A} A_{\mu_{\sigma(i,j)}}  
\fqr 
The integrals have to be performed in the former.  Contractions of 
the internal fields generate the propagators, and due to the tensor 
structure the integrals are more complicated.  However, all internal 
momenta or derivatives may be extracted from within the integral 
with the use of the identity, 

\bqr 
\partial^\mu {\triangle}^L = L \left[ (2-d) + m^2 \partial_m^2 \right] 
  \times ({k^\mu\over k^2}) {\triangle}^L \ .
\fqr
The massive propagator is 

\bqr
\triangle_{\mu\nu} = (x^2)^{-d/2+1} K_{d/2}(mx)  \quad  P_{\mu\nu} \ ,  
\fqr 
in terms of the modified Bessel function and the tensor structure (gauge 
dependent) of the propagator.  This identity is easily proved in x-space, 
and its application to the gauge field correlations is, upon iteration, 

\bqr
\int e^{ix\cdot k} \Prod_{j=1}^n \partial_{\mu_j} \triangle^L =
\sum_{\sigma,\sigma'} \prod \eta_{\mu_\sigma \nu_\sigma} 
\prod k_{\mu_\sigma'} ({1\over k^{2n_1+2n_2}}) (-2)^{n_2-1}  
\times 
  \left[ (2-d) + m^2 \partial_m^2 \right]^n \triangle^L  \ . 
\fqr
We have to sum over all pairs $(n_1,n_2)$ such that $n_1+n_2=n$ 
and all partitions of $n$ to the $\eta$ and $k$'s.  The gauge boson 
scattering is, 

\bqr  
\langle \left[ \Prod_{i=1}^L \left( \Prod_{j=1}^{m_i^\pa}
    \pa_{\mu_{\sigma'(i,j)}} \Prod_{j=1}^{m_i^A} A_{\mu_i}
        \right) \right] 
    \left[ \Prod_{i=1}^L \left( \Prod_{j=1}^{{\tilde m}_i^\pa}
  \pa_{\mu_{{\tilde\sigma}'(i,j)}} \Prod_{j=1}^{{\tilde m}_i^A}
        A_{{\tilde\mu}_i} \right) \right] 
\rangle 
\fqr 
\bqr 
= \sum_{\sigma,\sigma'} \prod \eta_{\mu_\sigma \nu_\sigma} 
\prod k_{\mu_\sigma'} ({1\over k^{2n_1+2n_2}}) (-2)^{n_2-1}  
\times 
  \left[ (2-d) + m^2 \partial_m^2 \right]^n \triangle^L  
\times \prod_{\rho(i),\tilde\rho(j)} 
\eta_{\mu_{\tilde\rho(i)}\nu_{\tilde \rho(j)}} \ .
\fqr 
The tensor structure follows from the deriavtive diagram, and the last 
term comes from the Pfaffian, 

\bqr 
\langle \prod_i^L A_{\mu_i} \prod_j^L A_{\tilde\mu_j} \rangle \ .
\fqr
The expression may seem complicated, but all loop integrals have been 
performed.  The algebraic complexity is also less than typical loop diagram 
calculations, which at intermediate steps may contain millions of terms.  

The integrals are evaluated to, 

\bqr 
\int d^dx \quad e^{ix\cdot k} \triangle^L = (k^2)^{-d/2-L(d/2-1)} 
  \sum_n (k^2/m^2)^n \alpha_n^{(L)} 
\fqr
in dimensional regularization, and 

\bqr 
(k^2)^{-d/2-L(d/2-1)} \sum_{m,n} ({k^2\over \Lambda^2})^m 
({k^2\over m^2})^n \alpha_{m,n}^{(L)}       
\fqr 
in a momentum cutoff scheme.  $N=4$ is finite, and divergences at 
intermediate steps in the calculation should drop out in the final equations. 

Equating the former with the quantum vertices generates a recursion in 
the couplings and tensor $t t \sim t$, with the coupling structure found from 
the 
ring of Eisenstein functions up to number coefficients.  The products of 
Eisenstein 
function iterate from $n$ and $m$-point to the higher derivative terms in $t$ 
(together with the enlarged basis in the latter).  The iteration formulae 
allow for 
a computation of the scattering amplitude via polynomial equations, with no 
integrals.  Similar results are available for the correlations of the 
composite 
operators. 

\section{Discussion}

The scattering amplitudes of spontaneously broken $N=4$ supersymmetric 
gauge theory have been examined in the derivative expanson.  The gauge 
theory may be iteratively constructed via sewing, and is equivalent to the 
sewing 
in the Feynman diagram expansion.   Unitarity is obvious in this 
construction. 
Amplitudes may be constructed without any integrals; the integrals involved 
are 
of the free-field type and may all be integrated.  Unitarity is obvious in 
this 
construction.

Scalar and gauge field theory has also been examined in \cite{scalar,gauge}; 
S-duality has been 
implemented in the $N=4$ gauge theory and order by order the gauge boson 
expansion is invariant under $SL(2,Z)$ duality.    S-duality in the 
amplitudes 
has the advantage that from the perturbative structure, all of the instanton 
corrections are computable and follow from the former.  The instanton 
expansion via the terms, $e^{-2n\tau_2}$,   
are connected to the perturbative terms and are redundant; the perturbative 
terms generate an instanton calculus via an $SL(2,Z)$ completion. 

The derivative expansion, and the concise expressions which follow from it, 
should enable further progress several areas.  The intermediate coupling 
regime is accessible as well as some possible information regarding the 
verification of duality.  

\section*{Acknowledgements} 

GC thanks the DOD, 444025-HL-25619, for support.

\end{document}